\documentclass[aip,apl,amsmath,amssymb,reprint,superscriptaddress,preprintnumbers,showpacs]{revtex4-1}
\pdfoutput=1
\usepackage{bm,mathrsfs}
\usepackage[mathcal]{euscript}
\usepackage[pdftex]{hyperref}
\usepackage{dcolumn}
\usepackage{graphicx}
\renewcommand{\vec}[1]{\bm{#1}}
\begin{document}

\title{Off-centered immobile magnetic vortex under influence of spin-transfer torque}

\author{Volodymyr P. Kravchuk}
 \email[Corresponding author. Electronic address: ]{vkravchuk@bitp.kiev.ua}
    \affiliation{Institute for Theoretical Physics, 03143 Kiev, Ukraine}

\author{Denis D. Sheka}
     \affiliation{Taras Shevchenko National University of Kiev, 01601 Kiev, Ukraine}
     \affiliation{Institute for Theoretical Physics, 03143 Kiev, Ukraine}

\author{Franz G.~Mertens}
    \affiliation{Physics Institute, University of Bayreuth, 95440 Bayreuth, Germany}

\author{Yuri Gaididei}
    \affiliation{Institute for Theoretical Physics, 03143 Kiev, Ukraine}

\date{\today}

%
%

\begin{abstract}
Equilibrium magnetization distribution of the vortex state magnetic
nanoparticle is affected by the influence of the spin-transfer
torque: an off-center out--of--plane vortex appears in the case of
the disk shape particle and pure planar vortex in the case of
asymmetric ring shape particle. The spin current causes extra
out-of-plane magnetization structures identical to well known dip
structures for the moving vortex. The shape of the dip structure
depends on the current strength and value of the off-center
displacement and it does not depend on the vortex polarity. The
critical current depends on the nanodot thickness.
\end{abstract}




\maketitle

The control of magnetic nonlinear structures using an electrical
current is of special interest for applications in
spintronics.\cite{Zutic04,*Tserkovnyak05} The spin--transfer torque
acts on nonhomogeneities in magnetization distributions, in
particular, on magnetic vortices.
\cite{Shibata06,Caputo07,*Sheka07b,Yamada07,Ivanov07b} In
particular, it was predicted theoretically \cite{Caputo07,*Sheka07b}
and confirmed experimentally \cite{Yamada07,Bolte08} that the vortex
core magnetization (so--called vortex polarity) can be switched on a
picosecond time scale. This discovery demonstrates the potential of
realizing all-electrically controlled magnetic memory devices,
changing the direction of the modern spintronics. \cite{Cowburn07}
Recently Shibata \textit{et al.}\cite{Shibata06} used the in-plane
spin current to demonstrate the effect of the spin-transfer torque
on the vortex state magnetic nanodisk. The spin current excites the
spiral motion of the vortex which finally relaxes to some shifted
position. Such a picture is a result of a Thiele--like vortex
dynamics, where the vortex does not change the shape during its
motion and it is valid for the relatively small vortex shifts.

In this Letter we predict the dip formation nearby the vortex under
the influence of the current. We consider two geometries, nanodisk
and asymmetric nanoring. The spin current in a latter case provides
the pure planar vortex centered on inner hole of the ring. Using
micromagnetic simulations \footnote[271]{We used the open source
\textsf{OOMMF} code \url{http://math.nist.gov/oommf/} with an
extension package for spin-current simulations developed by IBM
Zurich Research Laboratory
\url{http://www.zurich.ibm.com/st/magnetism/spintevolve.html}} we
discovered that such a shifted \textit{immobile} vortex gets an
extra out--of--plane magnetization which corresponds to the
well--known ``dip" structure of the moving vortex.
\cite{Waeyenberge06,*Hertel07,*Vansteenkiste09} Our study shows that
the dip development is specified by the current direction and its
intensity, and it does not depend on the vortex polarity. We also
show that the  critical current strongly depends on the nanodot
thickness.

\begin{figure}
\includegraphics[width=\columnwidth]{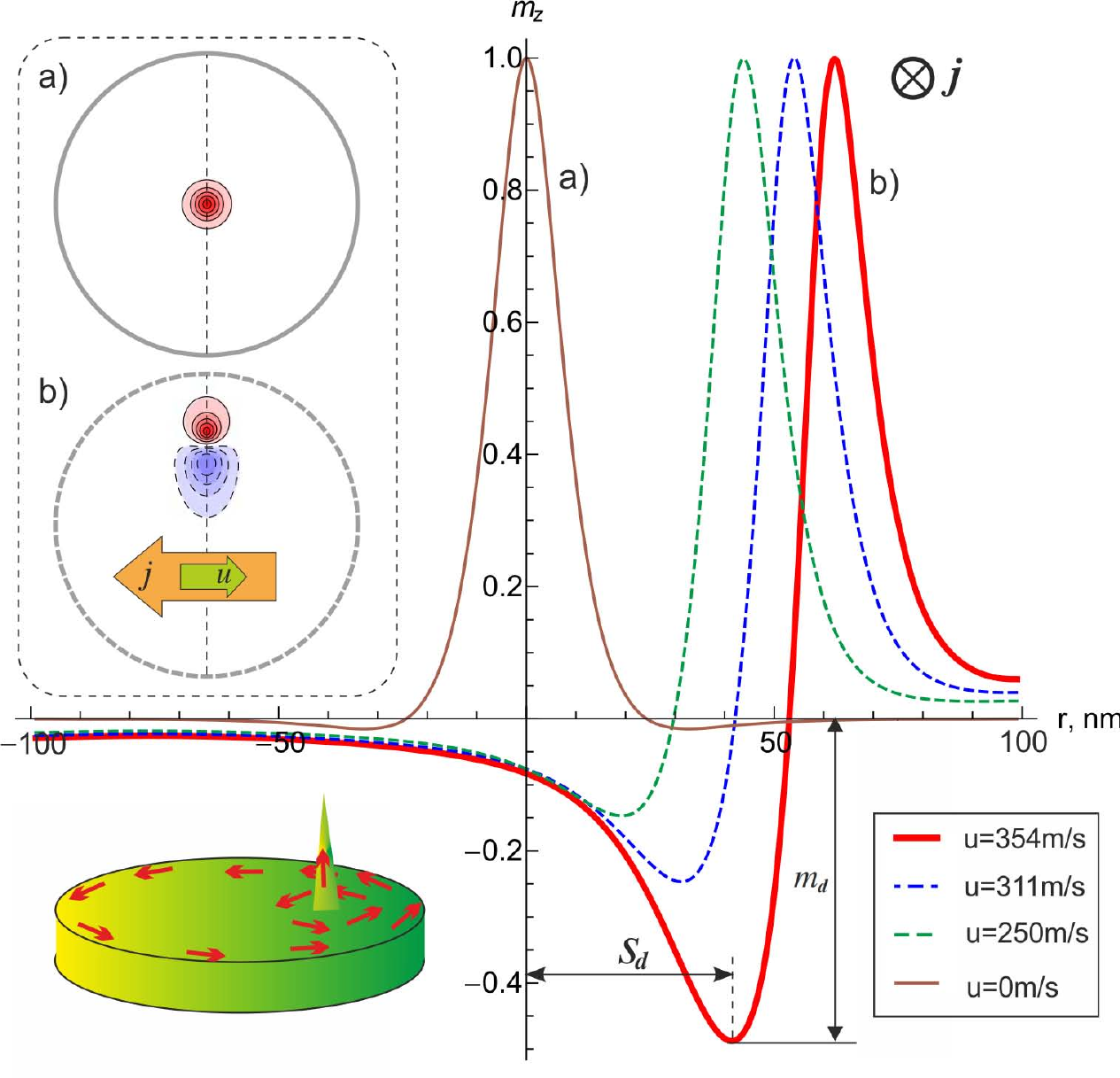}
\caption{ %
\label{fig:OPV} %
Profiles of the  \textit{immobile} equilibrium vortices for different values of the applied current. The profiles were taken along the diameter line perpendicular to the current direction -- dot-dashed line on the insets a) and b). The insets demonstrates the 2D distribution of the out-of-plane magnetization for two cases: zero current -- inset a) (the corresponding profile is shown with thin solid line), and  current close to critical one -- inset b) (the corresponding profile is shown with thick solid line). Dashed isolines corresponds to the case $m_z<0$ and solid isolines -- to $m_z>0$. All presented data were obtained from the micromagnetic simulations of the permalloy nanodisk with radius $L=100nm$ and thickness. $h=20nm$.} %
\end{figure}

\begin{figure}
\includegraphics[width=\columnwidth]{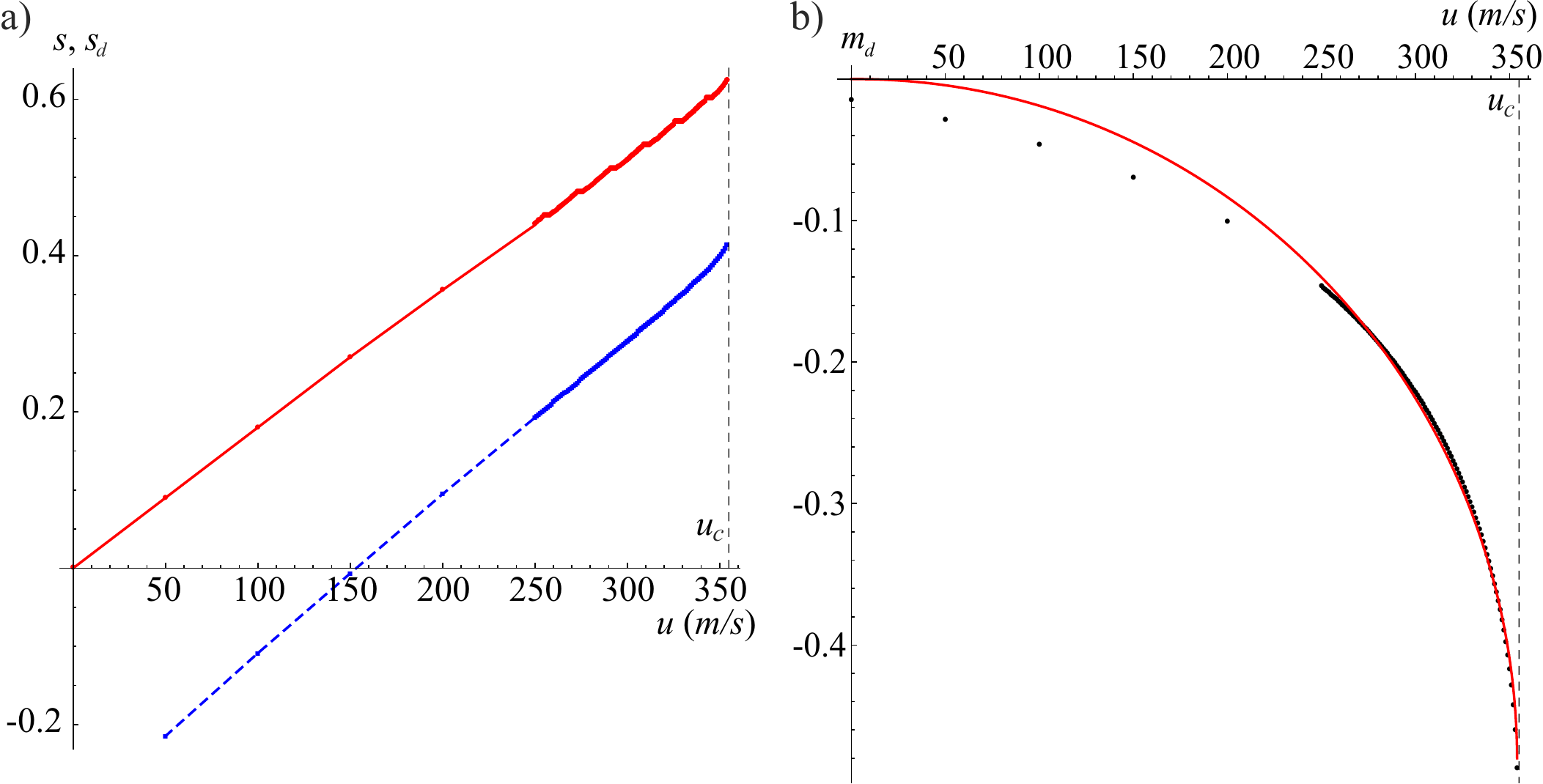}
\caption{%
\label{fig:params} %
Parameters of the equilibrium vortex state depending on the applied current. Plot a) shows the vortex (solid line) and dip (dashed line) displacement from the disk center depending on the velocity $u$. The displacements are measured in units of $L$. Plot b) shows dependence of the dip depth $m_d$ on $u$: points represents the simulation data and curve corresponds to Eq.~\eqref{eq:fit}. The vertical dashed line denotes the critical current $u_c$ when the vortex-antivortex pair is born.} %
\end{figure}

Our study is based on a simulated magnetization dynamics in the framework of the modified Landau-Lifshitz equation with the adiabatic spin torque term: \cite{Zhang04,*Thiaville05}
\begin{equation}\label{eq:LL}
\dot{\vec{m}} = -\gamma \vec{m}\times\vec{H}_{\text{eff}} + \alpha\vec{m}\times \dot{\vec{m}} - \left(\vec{u}\cdot \vec{\nabla}\right)\vec{m}.
\end{equation}
Here $\vec m$ is normalized magnetization vector: $\vec m=\vec M/M_{S}$ with $M_{S}$ being the saturation magnetization, $\gamma>0$ is giromagnetic ratio, $\vec{H}_{eff}$ is the effective micromagnetic field, $\alpha$ is the Gilbert damping constant. The velocity $\vec u$ is directed along the direction of electrons flow, with an amplitude $u\propto j$, where $j$ is the current density, which is supposed to be spatially uniform and constant. The permalloy \footnote[314]{In all \textsf{OOMMF} simulation we used material parameters adopted for permalloy: the exchange constant $A=1.3\times10^{-11}\mathrm{J/m}$, the saturation magnetization $M_S=8.6\times10^5\mathrm{A/m}$. This corresponds to the exchange length $\ell=\sqrt{A/4\pi M_S^2}\approx 5.3$ nm.} disk with radius $L=100nm$ and thickness $h=20nm$ was chosen for the numerical experiment. The ground state of such a disk is the out--of--plane vortex, situated at the disk origin [see Fig.~\ref{fig:OPV}a)]. Then the spin current velocity was switched on adiabatically: step--by--step we increased the value of velocity $u$ by a small value. On each step the full relaxation was achieved and the set of parameters was determined: position of the vortex center $s$, position of the dip minimum $s_d$ and also the dip depth $m_d$ (see Fig.~\ref{fig:OPV}). As one can see Fig.~\ref{fig:params}, the vortex and dip displacements linearly depend on the applied current. When the vortex shifts the dip amplitude increases and dependence $m_d(u)$ is essentially nonlinear, [see Fig.~\ref{fig:params}, inset b)]:
\begin{equation}\label{eq:fit}
m_d \propto
\begin{cases}
u,                        & \text{when $u\ll u_c$},\\
\sqrt{1-\dfrac{u}{u_c}}-1 & \text{when $u\lesssim u_c$}.
\end{cases}
\end{equation}

When the velocity $u$ achieves some critical value $u_c$ the dip
structure become unstable and depth abruptly achieves its minimum
value $m_d=-1$, hence the vortex-antivortex pair is born. Then the
vortex polarity switching occurs accordingly to the well studied
scenario.\cite{Waeyenberge06,Hertel07} It should be noted that the
obtained critical velocity $u_c=355 m/s$ is very close to the
critical velocity for moving vortex.\cite{Lee08c}

In order to study the role of the vortex out--of--plane structure in the dip formation process we performed the second kind of numerical experiment. With this end in view we considered the shifted vortex without out--of--plane component, the pure \emph{immobile planar} vortex. The vortex was pinned on the small hole placed in the half of the disk radius. The hole radius $r_h$ was chosen to be in the range $r_h\in (a_c;r_c\sim\ell)$, where $a_c$ is the critical radius of the transition between out-of-plane and planar vortices, \cite{Kravchuk07} $r_c$ is vortex core radius and $\ell$ is the exchange length \cite{Note314}. Thus the hole radius $r_h$ is big enough to make the vortex planar and it is small enough to prevent significant influence on the out--of--core magnetization structures. For thin permalloy films $a_c=0.37\ell\approx2$ nm \cite{Kravchuk07} and $\ell\approx5.3$ nm \cite{Note314}. That is why we chose $r_h=4$ nm. For simulation was chosen disk with $L=100$ nm and $h=10$ nm. The thickness decrease is needed for increasing the pinning effect due to decreasing the volume magnetostatic charges.

\begin{figure}
\includegraphics[width=\columnwidth]{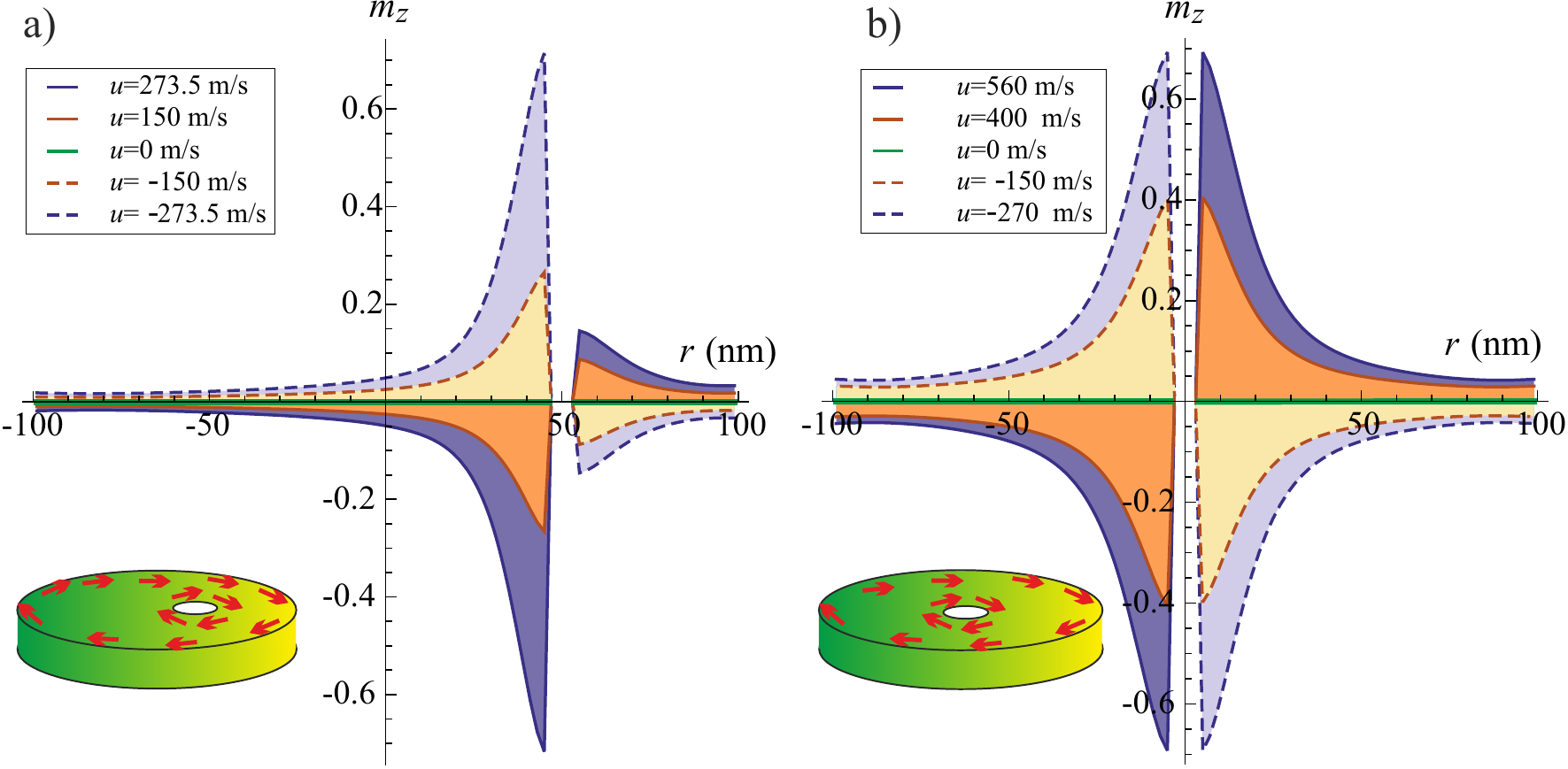}
\caption{%
\label{fig:shapes} %
In-plane vortex profiles for different current values and vortex positions. Dashed and solid lines corresponds to the opposite current directions. The inset a) shows profiles of the initially in-plane vortex pinned by the hole displaced by the value $L/2$ from the disk center. The inset b) shows the similar profiles for the centered vortex. Simulations were performed for the permalloy disk with $L=100$ nm and $h=10$ nm.} %
\end{figure}

Let us consider the planar vortex which is pinned at the inner hole of the ring. In the case of the symmetric hole, the planar vortex forms a ground state of the disk [see Fig.~\ref{fig:shapes}b)]. In the case of the asymmetric hole, the planar vortex is a equilibrium state, which corresponds to the local energy minimum in the current absence. One can see from the Fig.~\ref{fig:shapes} the vortices does not have core or any other out--of--plane components for the case of zero current. However under the influence of the current there appears a dip--like out--of--plane structure. This dip structure has the following main properties: (i) The dip structure of the centered vortex is symmetrical and its sign is changed when the sign of $u$ is changed, see Fig.~\ref{fig:shapes}b). (ii) The skewness appears when the vortex is shifted and its value rapidly increases when the displacement $s$ increases. (iii) The sign of the dip structure of the shifted vortex is determined by the sign of the product $su$ and it has no direct relation to the vortex polarity. (iv) The critical current $u_c$ takes maximal values for the centered vortex and decreases as the vortex displacement $s$ increases. The detailed analytical description of the dip development, based on the magnon mode analysis, is under consideration.

When the effective current velocity $u$ achieves its critical value $u_c$ the vortex-antivortex pair is born on the edge of the inner hole. The antivortex falls into the hole (it annihilates with the pinned vortex), so finally the new vortex appears out of the hole. Polarity of the new born vortex $p$ is equal to the polarity of the dip, in the other words $p=-\mathrm{sign}(us)$. In the case of centered vortex ($s=0$) the polarity $p$ is determined in a random way.

\begin{figure}
\includegraphics[width=\columnwidth]{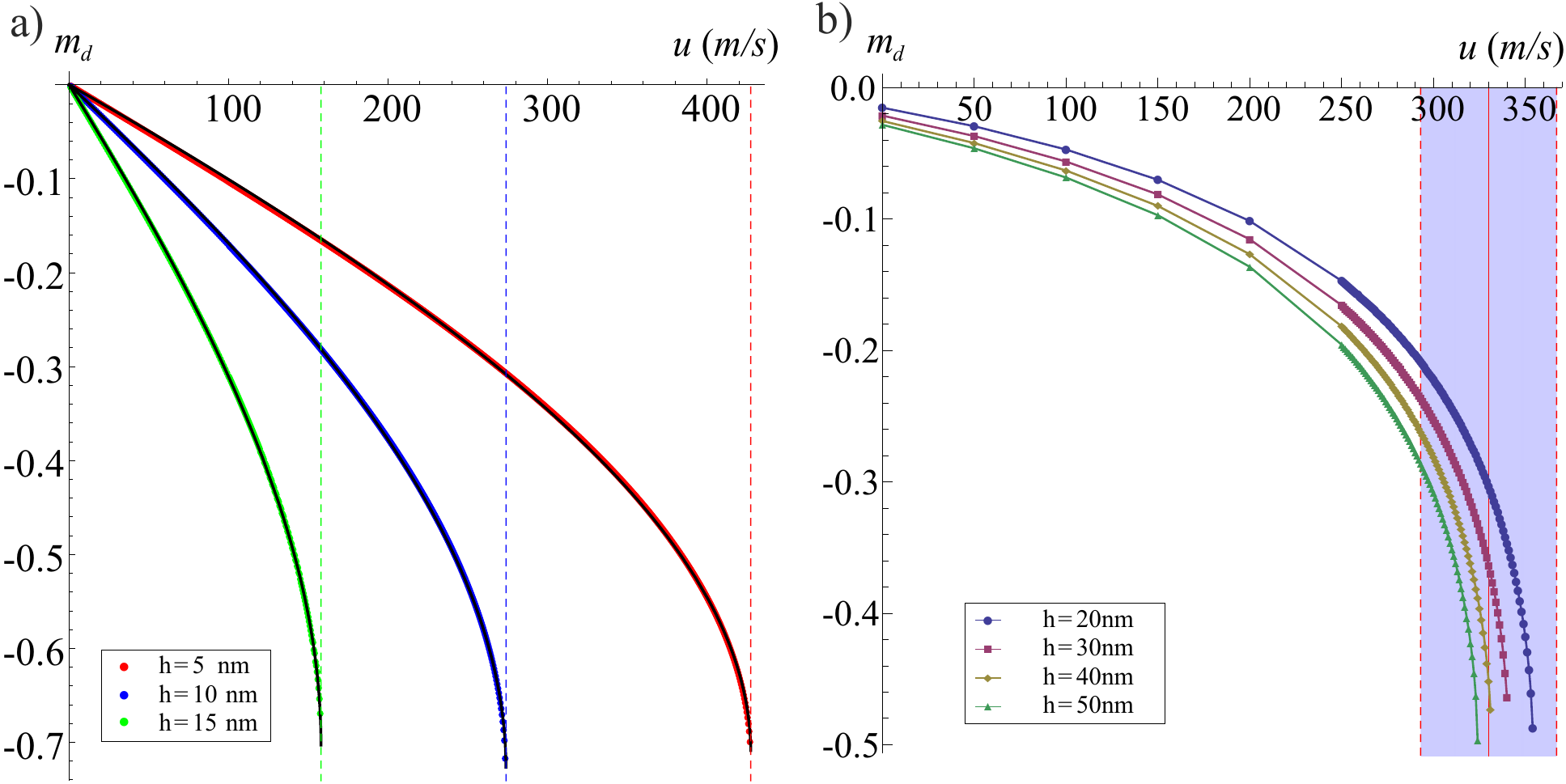}
\caption{%
\label{fig:dip_vs_u} %
Dependencies of the dip depth on the applied current for different thicknesses. Inset a) -- in-plane vortex is pinned on the hole with $s=0.5$, b) -- free out-of-plane vortex. Solid black lines demonstrates the dependence \eqref{eq:fit}. The critical velocity $u_c=330\pm37$ m/s which was determined in Ref.~\onlinecite{Lee08c} is shown in the subfigure b) as vertical strip.} %
\end{figure}

It is interesting to note that the critical current value depends on disk thickness and this dependence is different for free (out-of-plane) and pinned (planar) vortices, see Fig.~\ref{fig:dip_vs_u}. In the case of free vortex the critical velocity $u_c$ weakly dependents on thickness while in the case of pinned vortex this dependence is drastic.

In conclusion, we found that immobile planar vortex in a nanoring forms a dip structure under the action of adiabatic spin current under the threshold value. Above its value the vortex--antivortex pair can be nucleated, which can cause the vortex switching process.

\begin{acknowledgments}
The authors acknowledge support from Deutsches Zentrum f{\"u}r Luft- und Raumfart e.V., Internationales B{\"u}ro des BMBF in the frame of a bilateral scientific cooperation between Ukraine and Germany, project No.~UKR~08/001. D.D.S. thanks the University of Bayreuth, where a part of this work was performed, for kind hospitality and acknowledge the support from the Alexander von Humboldt Foundation.
\end{acknowledgments}

%

\end{document}